\documentclass[preprint2]{aastex}
\usepackage{color}

\usepackage{epsfig}

\shorttitle{Young stars in an old bulge}
\shortauthors{M Ness et al. }

\begin{document}


\title{Young stars in an old bulge: a natural outcome of internal evolution in the Milky Way}


\author{M. NESS\altaffilmark{1},  Victor. P. DEBATTISTA\altaffilmark{2}, T.BENSBY\altaffilmark{3}, S. FELTZING\altaffilmark{3}, R. RO\v SKAR\altaffilmark{4}, D.R. COLE\altaffilmark{2},  J. A. JOHNSON \altaffilmark{5,6}, K. FREEMAN\altaffilmark{7} }
\affil{$^1$Max-Planck-Institut f\"ur Astronomie, K\"onigstuhl 17, D-69117 Heidelberg, Germany}
\affil{$^2$ Jeremiah Horrocks Institute, University of Central Lancashire, Preston PR1 2HE, United Kingdom}
\affil{$^3$Lund Observatory, Department of Astronomy and Theoretical Physics, Box 43, SE-221\,00 Lund, Sweden.} 
\affil{$^4$Institute for Theoretical Physics, University of Z\"urich, Wintherthurerstrasse 190, Z\"urich, CH-8057, Switzerland}
\affil{$^5$ Department of Astronomy, Ohio State University, 140 W. 18th Avenue, Columbus, OH 43210, USA}
\affil{$^6$ Center for Cosmology and Astro-Particle Physics, Ohio State University, 191 West Woodruff Ave, Columbus, OH 43210 }
\affil{$^7$ Research School of Astronomy and Astrophysics, Mount Stromlo Observatory, Cotter Road, Weston Creek, ACT 2611, Australia}
\email{ness@mpia.de}

\begin{abstract}

  The center of our disk galaxy, the Milky Way, is dominated by a boxy/peanut-shaped bulge. Numerous studies of the bulge based on stellar photometry have concluded that the bulge stars are exclusively old. The perceived lack of young stars in the bulge strongly constrains its likely formation scenarios, providing evidence that the bulge is a unique population that formed early and separately from the disk. However, recent studies of individual bulge stars using the microlensing technique have reported that they span a range of ages, emphasizing that the bulge may not be a monolithic structure. In this letter we demonstrate that the presence of young stars that are located predominantly near the plane is expected for a bulge that has formed from the disk via dynamical instabilities. Using an N-body$+$SPH simulation of a disk galaxy forming out of gas cooling inside a dark matter halo and forming stars, we find a qualitative agreement between our model and the observations of young metal-rich stars in the bulge. We are also able to partially resolve the apparent contradiction in the literature between results that argue for a purely old bulge population and those which show a population comprised of a range in ages; \textit{the key is where to look.} \\
\end{abstract}


\section{INTRODUCTION}

The boxy/peanut-shaped bulge of the Milky Way was revealed in the star
counts from the Two Micron All Sky Survey (2MASS) \citep{lopez2005}
and more recently has been mapped in detail across the inner extent
using red clump stars \citep{wegg2013}. It has been proposed that
small boxy or peanut-shaped bulges form via internal evolution of the
disk \citep{combes1981}. As the disk becomes unstable, it forms a
rotating bar which buckles and heats the disk vertically
\citep{raha1991}. The orbits of the stars originally in the bar are
extended vertically into orbits which now define the
boxy/triaxial/peanut-shaped bulge. A number of properties of the bulge
of the Milky Way have recently been shown to be consistent with the
generic properties of N-body models that form a bulge via internal
evolution from the redistributed stars of the disk. These include the
kinematics of stars in the bulge
\citep{howard2008,kunder2012,ness2013b} and the X-shaped profile
\citep{mcwilliam2010,nataf2010}, which provide important observational
constraints on models of formation
\citep{shen2010,ness2012,li2012,gardner2013}. Another important
observational constraint on the formation of the bulge is stellar
ages.

Studies of the color-magnidtude diagrams of several fields in the
bulge have shown that the best isochrone fits support a purely old
($>10$\,Gyr) stellar population \citep[i.e.][]{ortolani1995,zoccali2003,sahu2006,clarkson2008,brown2010,valenti2013}. This has been interpreted as evidence for a
classical bulge population, formed rapidly at early times and before
the disk, via mergers or dissipational collapse processes
\citep{ortolani1995,zoccali2003}.  Recent studies interpret different
signatures of formation in the boxy/peanut morphology and the stellar
ages and metallicities to argue for a composite bulge. These studies
\citep[i.e][]{babusiaux2010, hill2011} conclude that dissipational
collapse formation has played an important role in the formation of
the bulge, in addition to the dynamical instability processes.
Classical bulges are easily formed in $\Lambda$CDM simulations, which
very well describe the universe on scales larger than 1Mpc. However,
the density distribution of the stars across the bulge implies that if
there is any component of the bulge that has formed via mergers, as
$\Lambda$CDM formation theory would predict, it is a relatively minor
population \citep{wegg2013}.  Additionally, although individual disk
galaxies with small bulges can be produced in $\Lambda$CDM simulations
by invoking feedback \citep{governato2008, scannapieco2008, brook2012,
  dm2012}, the merging and hierarchical processes generally produce
very large central bulges that are unlike the Milky Way. Given that
disk galaxies with no bulges or with relatively small boxy or
peanut-shaped bulges like the Milky Way are quite common
\citep{lutticke2004}, their formation route in the context of the
evolution of the universe is critical for our understanding of galaxy
formation. To constrain the formation processes of the Milky Way and
interpret the signatures of formation, including ages, comparisons to
models of individual galaxies are key.

\section{AGES OF INDIVIDUAL BULGE STARS}

\begin{figure}[h]
\centering
\includegraphics[scale=0.38]{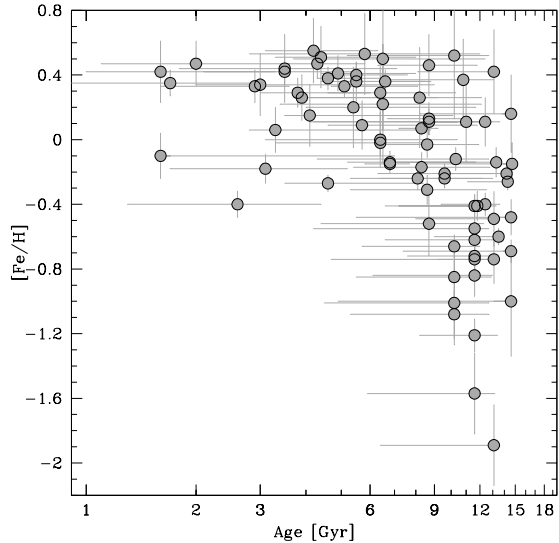}
\caption{Age-metallicity diagram of the 58 microlensed bulge dwarfs \citep{bensby2013} and an extra 20 dwarfs observed to date as part of their ongoing target of opportunity program. Note the metal-rich stars show a distribution in ages and metal-poor stare are all old. The
20 new observations will be published in Bensby et al. (2014, in prep.)}
\label{fig:bensby2}
\end{figure}

The old stellar ages reported from the photometric studies seem at
odds with the studies which exploit microlensing events of the dwarf
and subgiant bulge stars, which uniquely allow \textit{ages of
  individual} bulge stars to be determined. The microlensing studies
have demonstrated that the more metal-rich stars in the bulge
([Fe/H]$>$--0.4 dex) show a range in stellar ages from $3-12$\,Gyrs,
as shown in Figure \ref{fig:bensby2} \citep{bensby2011, bensby2013}.
Young stars are not expected to be present for a classical bulge
population, formed rapidly at early times and before the disk, via
dissipational or merger processes. If the bulge formed via internal
disk instabilities, rather than mergers, however, young stars should
be present due to ongoing star formation in the plane.

In this letter, we compare the age distribution of the microlensed
dwarf stars, which show a range of ages, to a high resolution
simulation that includes star formation, as described in Section 3. We
examine the stellar age distribution of the model as a function of
height from the plane and [Fe/H] in Section 4 and test the apparent
contradiction that exists between the range of ages reported from the
microlensing studies and the old population suggested by the
photometric studies.

\section{THE GALACTIC MODEL}

\begin{figure}[h]
\centering
\includegraphics[scale=0.22]{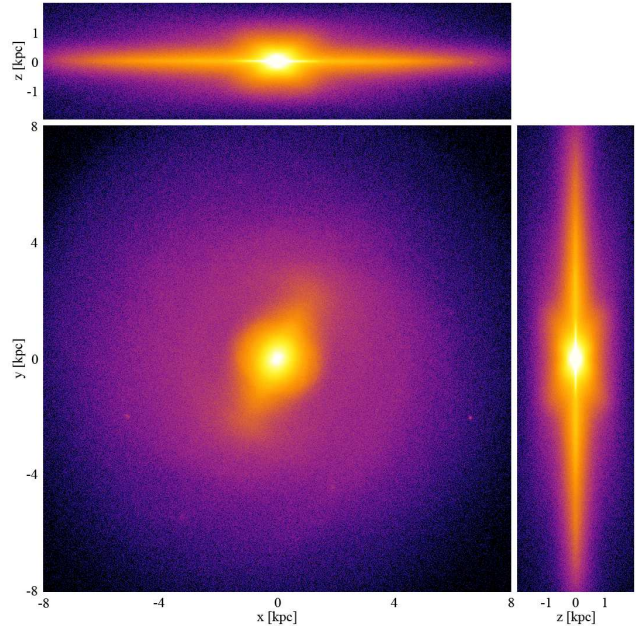}
\caption{Mass surface density of the model projected onto the $x-y$, $x-z$ and $y-z$ planes as indicated. Note the boxy shape and X-structure in the $y-z$ projection at right.  
  \label{fig:simulation}}
\end{figure}

Interpreting the observational data requires models that include star
formation. Until now, such predictions have come from semi-analytic
models or from cosmological simulations which generally do not yet
reach a resolution to resolve boxy/peanut-shaped structures.  Here we
use a self-consistent dissipational collapse simulation that is a
higher resolution version of previous models \citep{roskar2008a}, with
10$^{7}$ stars and a spatial resolution of 50 pc. In this model, a hot
gas corona in a Navarro Frenk White (NFW) \citep{navarro1997} dark
matter halo cools under self-gravity and forms a disk galaxy with a
small, weakly triaxial bulge. The evolution of this model in isolation
thus reflects internal processes only as there are no interacting
companions during the evolution.  The model was presented in
\citet{gardner2013}, the evolution of its nucleus will be presented in
Cole et al., 2014 (in preparation) and a more detailed analysis of the
evolution will be presented elsewhere (Ness et al., 2014).  For this paper
the model is studied after evolving for 10\,Gyr. We emphasise, this
model was not designed to match the Milky Way and our analysis is
qualitative.  However, we find a remarkable qualitative agreement with
a number of observational signatures. We have scaled the model (by a
factor of 1.2) to match the bar size to that of the Milky Way, of
about 3.5 kpc \citep{robin2012} and placed the bulge at 8 kpc away
from the Sun, at an angle of 27\,degrees with respect to the line of
sight \citep{wegg2013}. The model is shown in
Figure~\ref{fig:simulation} with the Sun placed at y = --8 kpc. It
qualitatively reproduces the observed kinematic profile that is
generic to models of bulges formed via internal evolution of the disk
\citep{shen2010,ness2013b} as well as the X-shaped profile that is
seen in the Milky Way \citep{gardner2013}.

\section{COMPARISON TO OBSERVATIONAL DATA}

\begin{figure*}
\centering
\includegraphics[scale=0.31]{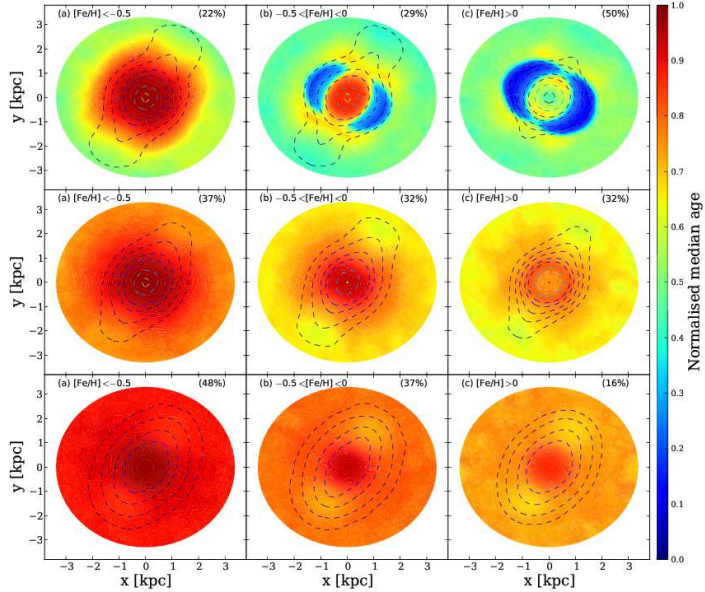}
\caption{The median age of stars in the inner  R$<$3 kpc of the simulation showing three slices in height, $|z|$, from the plane, from top to bottom of $|z|$$<$0.14 kpc, 0.14 kpc$<$$|z|$$<$0.42 kpc and 0.42 kpc$<$$|z|$$<$0.85 kpc. The color scale is the normalised median age. The sub panels represent stars of different metallicity bins: (a)  [Fe/H]$<$--0.5, (b) --0.5$<$[Fe/H]$<$ 0, (c) [Fe/H]$>$0. The density contours are shown to indicate the distribution of the stars in each bin. The percentage of stars in each metallicity bin is indicated in each sub-panel. }
\label{fig:ages2}
\end{figure*}

\begin{figure*}
\centering
\includegraphics[scale=0.31]{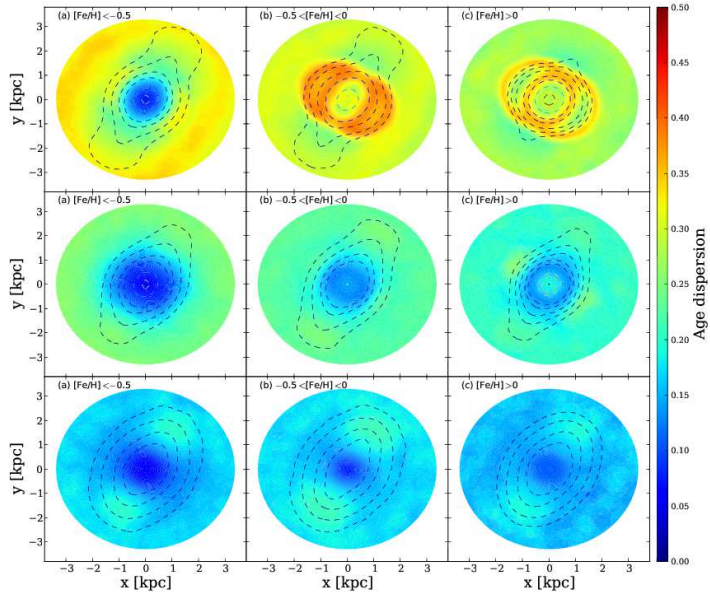}
\caption{The age dispersion of stars in the inner  R$<$3 kpc of the simulation with the same slices in $|z|$ and [Fe/H] bins as Fig 3.}
\label{fig:ages3}
\end{figure*}

The resolution of our model is much higher than in previous models,
and includes gas and stellar chemistries not previously employed in
such studies, so this is the first time structure can be studied in
such detail.  We wish to investigate the distribution of stellar ages
in the model at different heights from the plane and at different
metallicities, to understand if there are young stars in the bulge
and, if so, where they are concentrated.  Figures~\ref{fig:ages2}
and~\ref{fig:ages3} show the median age and age dispersion for the
face-on projections of the simulation out to a radius of {R $<$ 3
  kpc,} at three slices in height, $|z|$, from the plane. An age scale
normalised to the current age (=10 Gyr) of the simulation is used. The
surface density contours illustrate the extent of the bar.  These
$|z|$-slices correspond to latitudes of $|b|$$<$1$^\circ$,
1$^\circ$$<$$|b|$$<$3$^\circ$ and 3$^\circ$$<$$|b|$$<$6$^\circ$ from
the plane at the center of the bulge. At each height different
metallicity ranges are shown in the three sub-panels of: (a) [Fe/H]$<$---0.5, (b)
--0.5$<$[Fe/H]$<$--0, (c) [Fe/H]$>$0 and the percentage of stars is shown in the top right hand corner.

%

\subsection{THE AGE DISTRIBUTION OF STARS ACROSS LATITUDE}

Looking at Figure \ref{fig:ages2}, it is clear that according to the model (i)
the young stars are strongly concentrated very close to the plane,
(ii) there is a sharp transition between young and old stars in the
bulge with latitude and distinct transitions in the age distribution
with longitude (iii) the oldest stars become more centrally
concentrated at larger heights from the plane.  The model predicts
that a few degrees can make the difference between observing a
predominantly old population of stars or a population with a range of
stellar ages in the inner galaxy due to these steep gradients in the
age distribution in the bulge/bar and surrounding disk.  Nearest to
the plane around the inner region, the stars will have the largest age
range, so both young and old stars will be present. The regions around
the center of the bulge and ends of the bar show the most
homogeneously young population and the youngest stars are located
along the edges of the bar. Further away from the plane, the
inner-most bulge has the lowest dispersion in ages, reflecting a
predominantly old population of stars at the very center. At these
larger heights from the plane the younger stars are located
preferentially at the ends of the bar (as well as the surrounding
disk). The old population at the very center of the model has formed
early from a thick disk and the younger stars are from ongoing star
formation in the plane associated with the dynamical formation
processes. This recent star formation happens naturally without any
tuning.  The star formation history is partly driven by the gas inflow
to the center and related to how strong the bar is at any particular
time.  The star formation rate within the inner 300 pc is not
monotonically declining and during the periods when the bar
strengthens there is an increasing star formation rate.

According to these Figures, a median old stellar population is
expected to be observed for most fields in the bulge, located even a
few degrees from the plane. Although a somewhat younger population of
stars is present at the ends of the bar, including at
3$^\circ$$<$$|b|$$<$6$^\circ$, it is difficult to isolate this
population observationally for stars integrated along a given line of
sight. Only photometric observations which target the near-end of the
bar close to the plane ($l$ $\sim$ 12$^\circ$, $|b|$ $<$ 3$^\circ$),
should reveal a slightly younger mean population compared to the inner
longitudes. The photometric studies which report old isochrones as the
best fit models to fields at $(l,b)$ = (+1$^\circ$,--2.9$^\circ$)
\citep{ortolani1995}, (+0.3$^\circ$,--6.2$^\circ$)
\citep{zoccali2003}, (+1.25$^\circ$,--2.65$^\circ$) \citep{sahu2006}
and (+10.3$^\circ$,--4.2$^\circ$), (--6.8$^\circ$,4.7$^\circ$)
\citep{valenti2013} are consistent with the predictions of this model.

\subsection{THE AGE DISTRIBUTION OF STARS ACROSS [Fe/H]}

\begin{figure*}
\centering
\includegraphics[bb =0 160 700 400,clip, scale=0.8]{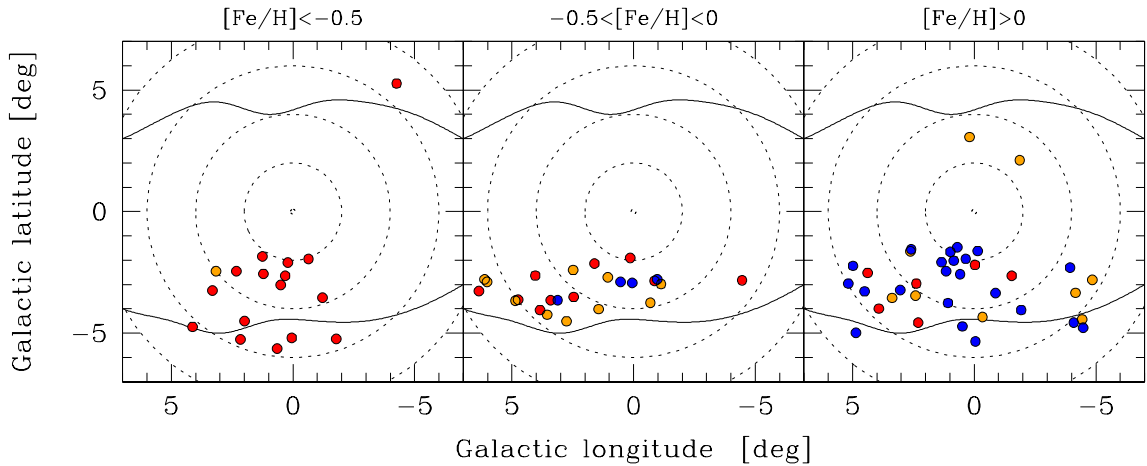}
\caption{The location on the sky of the 78 microlensed bulge dwarfs with red representing the oldest stars and blue the youngest with the metallicity intervals (a)-(c) as indicated on the top of each panel. The bulge contour lines based on observations with the COBE satellite are shown as solid lines \citep{weiland1994}. The dotted lines are concentric circles in steps of $2^{\circ}$.
\label{fig:glonglat}
}
\end{figure*}

To study the age distribution as a function of metallicity in the
model, we now examine the three metallicity ranges as shown in
Figures~\ref{fig:ages2}a-c and~\ref{fig:ages3}a-c, individually. There is clearly a
different age distribution for the most metal-rich versus the most
metal-poor stars. The model shows that at a given height from the
plane, the more metal-rich stars will be younger than the more
metal-poor stars and that there is an age gradient with distance from
the plane for the metal-rich stars. The age gradient is more gradual
at larger heights and the age distribution becomes increasingly older
and with a smaller age dispersion further away from the plane. For the metal-rich
stars in sub-panels (b) and (c), with [Fe/H]$>$--0.5, the predominantly
younger stars are concentrated to the plane and extend to the inner
most bulge at the highest metallicities. There is also a wide
dispersion in age nearest to the plane, where both young and old stars
will be observed.  It is likely, given the large fraction of young
metal-rich microlensed dwarf stars (see Figure \ref{fig:bensby2}),
that the youngest of these metal-rich stars are located in the ends of
the bar and inner disk directly surrounding the bulge/bar.

For the metal-poor stars, [Fe/H]$<$--0.5, shown in panel (a), the
model predicts that an old population is expected in the bulge, with younger stars located near the ends of the bar. Old stars of comparable ages to the bulge stars are also located in the nearby surrounding disk, and out to larger radii in the disk at $b$ $>$ 3$^\circ$.  This is an important
prediction of the model: a metal-rich star within $|b|$$<$6$^\circ$
from the plane in the bulge and the nearby surrounding disk can have a
range of ages, with preferentially younger stars at the highest
metallicities and lower latitudes, but a metal-poor star in the bulge,
regardless of latitude, will almost always be old.  Observationally,
the metallicity limit between these regimes in the Milky Way appears
around [Fe/H] $\approx$ --0.4 (see Figure \ref{fig:bensby2}).
Figure~\ref{fig:glonglat} shows the $(l,b)$ of the microlensed dwarf
sample, as a function of metallicity and colour coded by age where red
is the oldest stars and blue the youngest. This longitude range of the
microlensed stars of $|l|$ $<$ 6$^\circ$, in this Figure, corresponds
to $x$ = $\pm$ 0.5 kpc at a distance of 5 kpc from the Sun and $x$ =
$\pm$1.15 kpc at a distance of 11 kpc from the Sun, for comparison to
Figures \ref{fig:ages2} and \ref{fig:ages3}. Figure~\ref{fig:glonglat}
shows that the observed metal-poor stars are all old and present at
low and high latitudes and the dispersion in the ages of the stars
increases as a function of [Fe/H], in agreement with the distribution
seen in the model. From comparing the age distribution for all stars as a function of metallicity, it is clear that studies of individual stars
are necessary to test the range of ages of stars in the inner Galaxy.

\section{CONCLUSION}

Multiple channels of evidence including the kinematics
\citep{kunder2012, shen2010,ness2013a}, the X-shaped morphology
\citep{mcwilliam2010,nataf2010,ness2012,li2012} and now ages, as we
have demonstrated in this letter, indicate that the bulge of the Milky
Way is in large part, if not entirely, formed as part of the evolution
of the disk. This is at odds with the merger formation scenario
predicted by early semi-analytic models and a a significant problem
for the assumption that the bulge must have been accreted
\citep{guo2011,zavala2012,cooper2013}, given that Milky Way type
galaxies are common in the Universe. Our results demonstrate that
merger processes are not necessary to form the old bulge population
that has been previously associated with a classical bulge in
simulations \citep{dm2012,obreja2013} and early bulge-forming
dissipational collapse in the Milky Way \citep{zoccali2003,
  babusiaux2010, hill2011}. Our results are in agreement with the
bulge formation as a dynamical process from the disk at early times
reported by \citet{guedes2013} whereby the majority of stars are old.
In our analysis, we show that the presence of the young stars near the
plane in the simulations is an important aspect of bulge formation via
disk evolution, and one that matches well the observations
\citep{bensby2013}. Thus the observed ages of stars within the Milky
Way bulge can be explained within the context of internal formation
and dynamical processes, consistent with other lines of evidence
supporting a bulge largely forming from the disk. The key question we now endeavour to resolve, with respect to the Milky Way bulge as a
signature of the Galaxy's formation, is what fraction of stars, if
any, are part of a merger remnant? Mapping the age distribution of
the bulge as a function of $l$ and $b$ and performing further quantitative analysis which our team is preparing, examining key regions highlighted in this letter and further comparisons to models in the spirit of
this letter, is critical to interpret observations and understand the
formation of the Milky Way.

\section*{ACKNOWLEDGEMENTS}

The research leading to these results has received funding from the
European Research Council under the European Union's Seventh Framework
Programme (FP 7) ERC Grant Agreement n. [321035]. T.B. was funded by
grant No. 621-2009-3911 from The Swedish Research Council. V.P.D. and
D.R.C. are supported by STFC Consolidated grant \# ST/J001341/1. The
simulation used in this paper was run at the High Performance
Computing Facility of the University of Central Lancashire. We thank
Laurent Serge Noel of the University of Central Lancashire for Figure
2.


\begin{thebibliography}{38}

\bibitem[{{Babusiaux} {et~al.}(2010){Babusiaux}, {G{\'o}mez}, {Hill}, {Royer},
  {Zoccali}, {Arenou}, {Fux}, {Lecureur}, {Schultheis}, {Barbuy}, {Minniti}, \&
  {Ortolani}}]{babusiaux2010}
{Babusiaux}, C., {G{\'o}mez}, A., {Hill}, V., {Royer}, F., {Zoccali}, M.,
  {Arenou}, F., {Fux}, R., {Lecureur}, A., {Schultheis}, M., {Barbuy}, B.,
  {Minniti}, D., \& {Ortolani}, S. 2010, \aap, 519, A77

\bibitem[{{Bensby} {et~al.}(2011){Bensby}, {Ad{\'e}n}, {Mel{\'e}ndez}, {Gould},
  {Feltzing}, {Asplund}, {Johnson}, {Lucatello}, {Yee}, {Ram{\'{\i}}rez},
  {Cohen}, {Thompson}, {Bond}, {Gal-Yam}, {Han}, {Sumi}, {Suzuki}, {Wada},
  {Miyake}, {Furusawa}, {Ohmori}, {Saito}, {Tristram}, \&
  {Bennett}}]{bensby2011}
{Bensby}, T., {Ad{\'e}n}, D., {Mel{\'e}ndez}, J., {Gould}, A., {Feltzing}, S.,
  {Asplund}, M., {Johnson}, J.~A., {Lucatello}, S., {Yee}, J.~C.,
  {Ram{\'{\i}}rez}, I., {Cohen}, J.~G., {Thompson}, I., {Bond}, I.~A.,
  {Gal-Yam}, A., {Han}, C., {Sumi}, T., {Suzuki}, D., {Wada}, K., {Miyake}, N.,
  {Furusawa}, K., {Ohmori}, K., {Saito}, T., {Tristram}, P., \& {Bennett}, D.
  2011, \aap, 533, A134

\bibitem[{{Bensby} {et~al.}(2013){Bensby}, {Yee}, {Feltzing}, {Johnson},
  {Gould}, {Cohen}, {Asplund}, {Mel{\'e}ndez}, {Lucatello}, {Han}, {Thompson},
  {Gal-Yam}, {Udalski}, {Bennett}, {Bond}, {Kohei}, {Sumi}, {Suzuki}, {Suzuki},
  {Takino}, {Tristram}, {Yamai}, \& {Yonehara}}]{bensby2013}
{Bensby}, T., {Yee}, J.~C., {Feltzing}, S., {Johnson}, J.~A., {Gould}, A.,
  {Cohen}, J.~G., {Asplund}, M., {Mel{\'e}ndez}, J., {Lucatello}, S., {Han},
  C., {Thompson}, I., {Gal-Yam}, A., {Udalski}, A., {Bennett}, D.~P., {Bond},
  I.~A., {Kohei}, W., {Sumi}, T., {Suzuki}, D., {Suzuki}, K., {Takino}, S.,
  {Tristram}, P., {Yamai}, N., \& {Yonehara}, A. 2013, \aap, 549, A147

\bibitem[{{Brook} {et~al.}(2012){Brook}, {Stinson}, {Gibson}, {Ro{\v s}kar},
  {Wadsley}, \& {Quinn}}]{brook2012}
{Brook}, C.~B., {Stinson}, G., {Gibson}, B.~K., {Ro{\v s}kar}, R., {Wadsley},
  J., \& {Quinn}, T. 2012, \mnras, 419, 771

\bibitem[{{Brown} {et~al.}(2010){Brown}, {Sahu}, {Anderson}, {Tumlinson},
  {Valenti}, {Smith}, {Jeffery}, {Renzini}, {Zoccali}, {Ferguson},
  {VandenBerg}, {Bond}, {Casertano}, {Valenti}, {Minniti}, {Livio}, \&
  {Panagia}}]{brown2010}
{Brown}, T.~M., {Sahu}, K., {Anderson}, J., {Tumlinson}, J., {Valenti}, J.~A.,
  {Smith}, E., {Jeffery}, E.~J., {Renzini}, A., {Zoccali}, M., {Ferguson},
  H.~C., {VandenBerg}, D.~A., {Bond}, H.~E., {Casertano}, S., {Valenti}, E.,
  {Minniti}, D., {Livio}, M., \& {Panagia}, N. 2010, \apjl, 725, L19

\bibitem[{{Clarkson} {et~al.}(2008){Clarkson}, {Sahu}, {Anderson}, {Smith},
  {Brown}, {Rich}, {Casertano}, {Bond}, {Livio}, {Minniti}, {Panagia},
  {Renzini}, {Valenti}, \& {Zoccali}}]{clarkson2008}
{Clarkson}, W., {Sahu}, K., {Anderson}, J., {Smith}, T.~E., {Brown}, T.~M.,
  {Rich}, R.~M., {Casertano}, S., {Bond}, H.~E., {Livio}, M., {Minniti}, D.,
  {Panagia}, N., {Renzini}, A., {Valenti}, J., \& {Zoccali}, M. 2008, \apj,
  684, 1110

\bibitem[{{Combes} \& {Sanders}(1981)}]{combes1981}
{Combes}, F. \& {Sanders}, R.~H. 1981, \aap, 96, 164

\bibitem[{{Cooper} {et~al.}(2013){Cooper}, {D'Souza}, {Kauffmann}, {Wang},
  {Boylan-Kolchin}, {Guo}, {Frenk}, \& {White}}]{cooper2013}
{Cooper}, A.~P., {D'Souza}, R., {Kauffmann}, G., {Wang}, J., {Boylan-Kolchin},
  M., {Guo}, Q., {Frenk}, C.~S., \& {White}, S.~D.~M. 2013, \mnras, 434, 3348

\bibitem[{{Dom{\'e}nech-Moral} {et~al.}(2012){Dom{\'e}nech-Moral},
  {Mart{\'{\i}}nez-Serrano}, {Dom{\'{\i}}nguez-Tenreiro}, \& {Serna}}]{dm2012}
{Dom{\'e}nech-Moral}, M., {Mart{\'{\i}}nez-Serrano}, F.~J.,
  {Dom{\'{\i}}nguez-Tenreiro}, R., \& {Serna}, A. 2012, \mnras, 421, 2510

\bibitem[{{Gardner} {et~al.}(2013){Gardner}, {Debattista}, {Robin},
  {V{\'a}squez}, \& {Zoccali}}]{gardner2013}
{Gardner}, E., {Debattista}, V.~P., {Robin}, A.~C., {V{\'a}squez}, S., \&
  {Zoccali}, M. 2013, ArXiv e-prints

\bibitem[{{Governato} {et~al.}(2008){Governato}, {Mayer}, \&
  {Brook}}]{governato2008}
{Governato}, F., {Mayer}, L., \& {Brook}, C. 2008, in Astronomical Society of
  the Pacific Conference Series, Vol. 396, Formation and Evolution of Galaxy
  Disks, ed. J.~G. {Funes} \& E.~M. {Corsini}, 453

\bibitem[{{Guedes} {et~al.}(2013){Guedes}, {Mayer}, {Carollo}, \&
  {Madau}}]{guedes2013}
{Guedes}, J., {Mayer}, L., {Carollo}, M., \& {Madau}, P. 2013, \apj, 772, 36

\bibitem[{{Guo} {et~al.}(2011){Guo}, {White}, {Boylan-Kolchin}, {De Lucia},
  {Kauffmann}, {Lemson}, {Li}, {Springel}, \& {Weinmann}}]{guo2011}
{Guo}, Q., {White}, S., {Boylan-Kolchin}, M., {De Lucia}, G., {Kauffmann}, G.,
  {Lemson}, G., {Li}, C., {Springel}, V., \& {Weinmann}, S. 2011, \mnras, 413,
  101

\bibitem[{{Hill} {et~al.}(2011){Hill}, {Lecureur}, {G{\'o}mez}, {Zoccali},
  {Schultheis}, {Babusiaux}, {Royer}, {Barbuy}, {Arenou}, {Minniti}, \&
  {Ortolani}}]{hill2011}
{Hill}, V., {Lecureur}, A., {G{\'o}mez}, A., {Zoccali}, M., {Schultheis}, M.,
  {Babusiaux}, C., {Royer}, F., {Barbuy}, B., {Arenou}, F., {Minniti}, D., \&
  {Ortolani}, S. 2011, \aap, 534, A80

\bibitem[{{Howard} {et~al.}(2008){Howard}, {Rich}, {Reitzel}, {Koch}, {De
  Propris}, \& {Zhao}}]{howard2008}
{Howard}, C.~D., {Rich}, R.~M., {Reitzel}, D.~B., {Koch}, A., {De Propris}, R.,
  \& {Zhao}, H. 2008, \apj, 688, 1060

\bibitem[{{Kunder} {et~al.}(2012){Kunder}, {Koch}, {Rich}, {de Propris},
  {Howard}, {Stubbs}, {Johnson}, {Shen}, {Wang}, {Robin}, {Kormendy}, {Soto},
  {Frinchaboy}, {Reitzel}, {Zhao}, \& {Origlia}}]{kunder2012}
{Kunder}, A., {Koch}, A., {Rich}, R.~M., {de Propris}, R., {Howard}, C.~D.,
  {Stubbs}, S.~A., {Johnson}, C.~I., {Shen}, J., {Wang}, Y., {Robin}, A.~C.,
  {Kormendy}, J., {Soto}, M., {Frinchaboy}, P., {Reitzel}, D.~B., {Zhao}, H.,
  \& {Origlia}, L. 2012, \aj, 143, 57

\bibitem[{{Li} \& {Shen}(2012)}]{li2012}
{Li}, Z.-Y. \& {Shen}, J. 2012, \apjl, 757, L7

\bibitem[{{L{\'o}pez-Corredoira} {et~al.}(2005){L{\'o}pez-Corredoira},
  {Cabrera-Lavers}, \& {Gerhard}}]{lopez2005}
{L{\'o}pez-Corredoira}, M., {Cabrera-Lavers}, A., \& {Gerhard}, O.~E. 2005,
  \aap, 439, 107

\bibitem[{{L{\"u}tticke} {et~al.}(2004){L{\"u}tticke}, {Pohlen}, \&
  {Dettmar}}]{lutticke2004}
{L{\"u}tticke}, R., {Pohlen}, M., \& {Dettmar}, R.-J. 2004, aap, 417, 527

\bibitem[{{McWilliam} \& {Zoccali}(2010)}]{mcwilliam2010}
{McWilliam}, A. \& {Zoccali}, M. 2010, \apj, 724, 1491

\bibitem[{{Nataf} {et~al.}(2010){Nataf}, {Udalski}, {Gould}, {Fouqu{\'e}}, \&
  {Stanek}}]{nataf2010}
{Nataf}, D.~M., {Udalski}, A., {Gould}, A., {Fouqu{\'e}}, P., \& {Stanek},
  K.~Z. 2010, \apjl, 721, L28

\bibitem[{{Navarro} {et~al.}(1997){Navarro}, {Frenk}, \& {White}}]{navarro1997}
{Navarro}, J.~F., {Frenk}, C.~S., \& {White}, S.~D.~M. 1997, \apj, 490, 493

\bibitem[{{Ness} {et~al.}(2013{\natexlab{a}}){Ness}, {Freeman}, {Athanassoula},
  {Wylie-de-Boer}, {Bland-Hawthorn}, {Asplund}, {Lewis}, {Yong}, {Lane}, \&
  {Kiss}}]{ness2013a}
{Ness}, M., {Freeman}, K., {Athanassoula}, E., {Wylie-de-Boer}, E.,
  {Bland-Hawthorn}, J., {Asplund}, M., {Lewis}, G.~F., {Yong}, D., {Lane},
  R.~R., \& {Kiss}, L.~L. 2013{\natexlab{a}}, \mnras, 430, 836

\bibitem[{{Ness} {et~al.}(2013{\natexlab{b}}){Ness}, {Freeman}, {Athanassoula},
  {Wylie-de-Boer}, {Bland-Hawthorn}, {Asplund}, {Lewis}, {Yong}, {Lane},
  {Kiss}, \& {Ibata}}]{ness2013b}
{Ness}, M., {Freeman}, K., {Athanassoula}, E., {Wylie-de-Boer}, E.,
  {Bland-Hawthorn}, J., {Asplund}, M., {Lewis}, G.~F., {Yong}, D., {Lane},
  R.~R., {Kiss}, L.~L., \& {Ibata}, R. 2013{\natexlab{b}}, \mnras, 432, 2092

\bibitem[{{Ness} {et~al.}(2012){Ness}, {Freeman}, {Athanassoula},
  {Wylie-De-Boer}, {Bland-Hawthorn}, {Lewis}, {Yong}, {Asplund}, {Lane},
  {Kiss}, \& {Ibata}}]{ness2012}
{Ness}, M., {Freeman}, K., {Athanassoula}, E., {Wylie-De-Boer}, E.,
  {Bland-Hawthorn}, J., {Lewis}, G.~F., {Yong}, D., {Asplund}, M., {Lane},
  R.~R., {Kiss}, L.~L., \& {Ibata}, R. 2012, \apj, 756, 22

\bibitem[{{Obreja} {et~al.}(2013){Obreja}, {Dom{\'{\i}}nguez-Tenreiro},
  {Brook}, {Mart{\'{\i}}nez-Serrano}, {Dom{\'e}nech-Moral}, {Serna},
  {Moll{\'a}}, \& {Stinson}}]{obreja2013}
{Obreja}, A., {Dom{\'{\i}}nguez-Tenreiro}, R., {Brook}, C.,
  {Mart{\'{\i}}nez-Serrano}, F.~J., {Dom{\'e}nech-Moral}, M., {Serna}, A.,
  {Moll{\'a}}, M., \& {Stinson}, G. 2013, \apj, 763, 26

\bibitem[{{Ortolani} {et~al.}(1995){Ortolani}, {Renzini}, {Gilmozzi},
  {Marconi}, {Barbuy}, {Bica}, \& {Rich}}]{ortolani1995}
{Ortolani}, S., {Renzini}, A., {Gilmozzi}, R., {Marconi}, G., {Barbuy}, B.,
  {Bica}, E., \& {Rich}, R.~M. 1995, \nat, 377, 701

\bibitem[{{Raha} {et~al.}(1991){Raha}, {Sellwood}, {James}, \&
  {Kahn}}]{raha1991}
{Raha}, N., {Sellwood}, J.~A., {James}, R.~A., \& {Kahn}, F.~D. 1991, \nat,
  352, 411

\bibitem[{{Robin} {et~al.}(2012){Robin}, {Luri}, {Reyl{\'e}}, {Isasi}, {Grux},
  {Blanco-Cuaresma}, {Arenou}, {Babusiaux}, {Belcheva}, {Drimmel}, {Jordi},
  {Krone-Martins}, {Masana}, {Mauduit}, {Mignard}, {Mowlavi},
  {Rocca-Volmerange}, {Sartoretti}, {Slezak}, \& {Sozzetti}}]{robin2012}
{Robin}, A.~C., {Luri}, X., {Reyl{\'e}}, C., {Isasi}, Y., {Grux}, E.,
  {Blanco-Cuaresma}, S., {Arenou}, F., {Babusiaux}, C., {Belcheva}, M.,
  {Drimmel}, R., {Jordi}, C., {Krone-Martins}, A., {Masana}, E., {Mauduit},
  J.~C., {Mignard}, F., {Mowlavi}, N., {Rocca-Volmerange}, B., {Sartoretti},
  P., {Slezak}, E., \& {Sozzetti}, A. 2012, \aap, 543, A100

\bibitem[{{Ro{\v s}kar} {et~al.}(2008){Ro{\v s}kar}, {Debattista}, {Stinson},
  {Quinn}, {Kaufmann}, \& {Wadsley}}]{roskar2008a}
{Ro{\v s}kar}, R., {Debattista}, V.~P., {Stinson}, G.~S., {Quinn}, T.~R.,
  {Kaufmann}, T., \& {Wadsley}, J. 2008, \apjl, 675, L65

\bibitem[{{Sahu} {et~al.}(2006){Sahu}, {Casertano}, {Bond}, {Valenti}, {Ed
  Smith}, {Minniti}, {Zoccali}, {Livio}, {Panagia}, {Piskunov}, {Brown},
  {Brown}, {Renzini}, {Rich}, {Clarkson}, \& {Lubow}}]{sahu2006}
{Sahu}, K.~C., {Casertano}, S., {Bond}, H.~E., {Valenti}, J., {Ed Smith}, T.,
  {Minniti}, D., {Zoccali}, M., {Livio}, M., {Panagia}, N., {Piskunov}, N.,
  {Brown}, T.~M., {Brown}, T., {Renzini}, A., {Rich}, R.~M., {Clarkson}, W., \&
  {Lubow}, S. 2006, \nat, 443, 534

\bibitem[{{Scannapieco} {et~al.}(2008){Scannapieco}, {Tissera}, {White}, \&
  {Springel}}]{scannapieco2008}
{Scannapieco}, C., {Tissera}, P.~B., {White}, S.~D.~M., \& {Springel}, V. 2008,
  \mnras, 389, 1137

\bibitem[{{Shen} {et~al.}(2010){Shen}, {Rich}, {Kormendy}, {Howard}, {De
  Propris}, \& {Kunder}}]{shen2010}
{Shen}, J., {Rich}, R.~M., {Kormendy}, J., {Howard}, C.~D., {De Propris}, R.,
  \& {Kunder}, A. 2010, \apjl, 720, L72

\bibitem[{{Valenti} {et~al.}(2013){Valenti}, {Zoccali}, {Renzini}, {Brown},
  {Gonzalez}, {Minniti}, {Debattista}, \& {Mayer}}]{valenti2013}
{Valenti}, E., {Zoccali}, M., {Renzini}, A., {Brown}, T.~M., {Gonzalez}, O.,
  {Minniti}, D., {Debattista}, V.~P., \& {Mayer}, L. 2013, ArXiv e-prints

\bibitem[{{Wegg} \& {Gerhard}(2013)}]{wegg2013}
{Wegg}, C. \& {Gerhard}, O. 2013, ArXiv e-prints

\bibitem[{{Weiland} {et~al.}(1994){Weiland}, {Arendt}, {Berriman}, {Dwek},
  {Freudenreich}, {Hauser}, {Kelsall}, {Lisse}, {Mitra}, {Moseley}, {Odegard},
  {Silverberg}, {Sodroski}, {Spiesman}, \& {Stemwedel}}]{weiland1994}
{Weiland}, J.~L., {Arendt}, R.~G., {Berriman}, G.~B., {Dwek}, E.,
  {Freudenreich}, H.~T., {Hauser}, M.~G., {Kelsall}, T., {Lisse}, C.~M.,
  {Mitra}, M., {Moseley}, S.~H., {Odegard}, N.~P., {Silverberg}, R.~F.,
  {Sodroski}, T.~J., {Spiesman}, W.~J., \& {Stemwedel}, S.~W. 1994, \apjl, 425,
  L81

\bibitem[{{Zavala} {et~al.}(2012){Zavala}, {Avila-Reese}, {Firmani}, \&
  {Boylan-Kolchin}}]{zavala2012}
{Zavala}, J., {Avila-Reese}, V., {Firmani}, C., \& {Boylan-Kolchin}, M. 2012,
  \mnras, 427, 1503

\bibitem[{{Zoccali} {et~al.}(2003){Zoccali}, {Renzini}, {Ortolani}, {Greggio},
  {Saviane}, {Cassisi}, {Rejkuba}, {Barbuy}, {Rich}, \& {Bica}}]{zoccali2003}
{Zoccali}, M., {Renzini}, A., {Ortolani}, S., {Greggio}, L., {Saviane}, I.,
  {Cassisi}, S., {Rejkuba}, M., {Barbuy}, B., {Rich}, R.~M., \& {Bica}, E.
  2003, \aap, 399, 931

\end{thebibliography}
\end{document}